\documentclass[twocolumn]{aastex631}

\usepackage{hyperref}
\usepackage{xcolor}
\usepackage{soul}
\usepackage{amsmath}

\graphicspath{{./}{figures/}}

\begin{document}
\title{Rotational flows in solar coronal flux rope cavities}

\correspondingauthor{Valeriia Liakh}
\email{valeriia.liakh@kuleuven.be}

\author[0000-0002-9570-8145]{Valeriia Liakh}
\affiliation{Centre for mathematical Plasma Astrophysics, Department of Mathematics,\\
	 KU Leuven, Celestijnenlaan 200B, 3001 Leuven, Belgium}

\author[0000-0003-3544-2733]{Rony Keppens}
\affiliation{Centre for mathematical Plasma Astrophysics, Department of Mathematics,\\
	KU Leuven, Celestijnenlaan 200B, 3001 Leuven, Belgium}

\newcommand{\Mm}{{\mathrm{\, Mm}}}
\newcommand{\kms}{{\mathrm{\, km \,s^{-1}}}}
\newcommand{\hours}{{\mathrm{\, hours}}}
\newcommand{\secs}{{\mathrm{\, seconds}}}
\newcommand{\mins}{{\mathrm{\, minutes}}}

\begin{abstract}
	
  We present a 2.5-dimensional magnetohydrodynamic simulation of a systematically rotating prominence inside its coronal cavity using the open-source \texttt{MPI-AMRVAC} code. Our simulation starts from a non-adiabatic, gravitationally stratified corona, permeated with a sheared arcade magnetic structure. The flux rope (FR) is formed through converging and shearing footpoints driving, simultaneously applying randomized heating at the bottom. The latter induces a left-right asymmetry of temperature and density distributions with respect to the polarity-inversion line. This asymmetry drives flows along the loops before the FR formation, which gets converted to net rotational motions upon reconnection of the field lines. As the thermal instability within the FR develops, angular momentum conservation about its axis leads to a systematic rotation of both hot coronal and cold condensed plasma. The initial rotational velocity exceeds $60\kms$. The synthesized images confirm the simultaneous rotations of the coronal plasma seen in 211 and 193 \AA\  and condensations seen in 304 \AA. Furthermore, the formation of the dark cavity is evident in 211 and 193 \AA\ images. Our numerical experiment is inspired by observations of so-called giant solar prominence tornadoes, and reveals that asymmetric FR formation can be crucial in triggering rotational motions. We reproduce observed spinning motions inside the coronal cavity, augmenting our understanding of the complex dynamics of rotating prominences.

\end{abstract}

\section{Introduction}

Prominences are clouds of cold and dense plasma that float in the hot and tenuous corona, supported against gravity by the complex magnetic field. The most plausible magnetic field structure supporting prominences are flux rope (FR) and sheared arcade (SA) \citep{Mackay:2010ssr}. Cavities observed as circular or semicircular dark regions at the limb in white light, soft X-ray, and extreme ultraviolet (EUV) observations \citep{Waldmeier:1970solphys,Saito:1973solphys,Serio:1978solphys,Gibson:2010apj,Reeves:2012apj} are believed to be the cross-sections of FRs or SAs. We refer to the review by \citet{Gibson:2015} for a more detailed discussion on solar cavities.

Prominences are highly dynamic structures that exhibit a variety of systematic flows, including rotations, that attracted great interest in the last decades. These spinning prominence motions were first described in detail and labeled as giant tornadoes by \cite{Pettit:1925}. 

Since then, the number of reported rotational events increased, with a number of these events observed in the prominence feet \citep{Su:2012apjl,Su:2014apjl,Orozco:2012apjl,Yan:2013aj,Yan:2014apj,Wedemeyer:2013apj,Wedemeyer:2014pasj} and inside the cavity \citep{Ohman:1969solphys,Liggett:1984solphys,Schmit:2009apjl,Wang:2010apjl,Li:2012apjl,Mishra:2020solphys}. The estimated velocities of these giant tornadoes vary in the range of $5-75\kms$.

Cavity tornadoes are believed to be plasma moving along a helical FR, which when viewed along the helix axis appears as a rotation \citep{Li:2012apjl,Panasenco:2014solphys}. In this way, the tornado can eventually disintegrate when material drains down one of the FR footpoints \citep{Wang:2017apj}. \citet{Panesar:2013aap} additionally analyzed the event reported by \citet{Li:2012apjl} and found that nearby solar flares resulted in changes to the magnetic energy, allowing the cavity to expand. The tornado was interpreted as the dynamic response to this expansion. \citet{Wang:2010apjl} suggested an alternative driving mechanism of the rotational motions inside the dark cavity. Asymmetric heating at the footpoints of the loop arcade structure can drive a flow. As the footpoints of the loops reconnect, forming a helical FR, the motion is converted into rotation in the same direction as the initial flow. \citet{Wang:2010apjl} suggested that there may be a preference for equatorward rotation in tornado-like events. \citet{Li:2012apjl} pointed out that this implies a preferential direction for the helicity of prominences and a preferential direction for material flow within the structure.

In this Letter, we demonstrate for the first time the self-consistent development of swirling plasma motions inside the cavity using a 2.5D magnetohydrodynamic (MHD) simulation of FR formation in the non-adiabatic gravitationally stratified corona under the presence of turbulent heating at the footpoints. This randomized heating has been used previously to study counterstreaming flows in sheared arcade solar prominences \citep{Zhou:2020nat,Jercic:2023aap} and coronal rain in magnetic arcade configurations \citep{Li:2022apj}, and is meant to mimick the unavoidable stochastic buffeting from convective layers below our simulated domain. This Letter is organized as follows: in Section \ref{sec:numerical-model}, the numerical setup is described; in Section \ref{sec:results}, we explain and discuss the evolution and main properties of the rotational flows inside FR, and in Section \ref{sec:summary}, the main results are summarized.

\section{Numerical model}\label{sec:numerical-model}

 The numerical experiment is performed using the fully open-source, adaptive-grid, parallelized Adaptive Mesh Refinement Versatile Advection Code\footnote{\texttt{MPI-AMRVAC 3.0} with website \href{http://amrvac.org}{http://amrvac.org}.} \citep{Porth:2014apjs,Xia:2018apjs,Keppens:2021cmwa,amrvac2023}. 
 We use a Cartesian coordinate system with $x$- and $y$-axis denoting the horizontal and vertical direction, respectively. The numerical domain has a physical size of $48\times 144$ Mm consisting of $96\times 288$ grid cells for the base resolution. We apply five levels of adaptive mesh so that we resolve structures down to our smallest grid cell size of $31.25$ km. We adopt a Lohner-type prescription \citep{Lohner:1987} for the refinement based on second order gradient evaluations of the density and magnetic field components.

 The \texttt{MPI-AMRVAC} code solves MHD equations that take into account non-ideal, non-adiabatic effects, and various physical source terms like the solar gravitational field. These standard MHD equations can be found in many textbooks \citep[e.g.][]{bookhans} and are equivalent to the Eqs. 2-5 presented in \citet{Brughmans:2022aap}. For the equation of state, we use the ideal gas law for a monoatomic gas with a specific heat ratio $\gamma=5/3$. The mean molecular mass is $\mu\approx0.6$, since we assume fully-ionized plasma with the Helium abundance $n_{He}=0.1n_{H}$. The temperature equation contains important terms, including optically thin radiation, anisotropic thermal conduction, and Ohmic heating due to a uniform and explicitly resolved magnetic resistivity, which is chosen at $\eta=2\times10^{-5}$ in code units. The energy balance equation also includes the background heating term in the form of exponential decay to balance radiative losses.

The set of MHD equations is solved using the finite volume scheme setup combining the total variation diminishing Lax-Friedrichs scheme TVDLF \citep{Yee:1989,Toth:1996} with the second-order symmetric TVD slope limiter \citep{Vanleer:1974}. We also perform another numerical experiment with the HLL flux scheme \citep{Harten:1983}. We note only minor differences in evolution, and the most important details are preserved.
The time integration is performed using the strong stability-preserving Runge-Kutta three-step third-order method (SSPRK3). 
In order to control $\nabla\cdot\mathbf{B}$, we use the parabolic diffusion method \citep{Keppens:2003,amrvac2023}. 
We also exploit the magnetic field splitting \citep{Xia:2018apjs}, fixing the equilibrium magnetic field $\mathbf{B}_0$ and separating the time-dependent perturbation $\mathbf{B}_1$.
For the thin radiative losses, we use \textsc{Colgan-DM} cooling curve from \citet{Colgan:2008apj}, but extended with a low temperature treatment, for details see \citet{Hermans2021}.  

The initial atmosphere is a gravitationally stratified corona assuming the constant temperature $T_0=1$ MK and the gravitational acceleration defined as $g(y)=g_{\odot}R_{\odot}^2/(R_{\odot}+y)^2$, where $g_{\odot}=2.74\times 10^{4} \mathrm{cm\ s^{-2}}$ is the gravitational acceleration at the solar surface and $R_{\odot}=695.7\Mm$ is the solar radius. The pressure scale height then varies with height and is defined as $H(y)=H_{0}(R_{\odot}+y)/R_{\odot}$, where $H_{0}\approx50$ Mm is the pressure scale height at the bottom. The pressure, density and background heating have values $p_{0,bot}=0.434\ \mathrm{dyn\ cm^{-2}}$, $\rho_{0,bot}=3.2\times10^{-15}\ \mathrm{g\ cm^{-3}}$, and $\mathcal{H}_b(y)=\rho_0^2\Lambda(T_0)e^{-2y/H(y)}= 10^{-3}\ \mathrm{ergs\ cm^{-3}\ s^{-1}}$ at $y=0$, respectively. We include anisotropic thermal conduction $\nabla\cdot(\overleftrightarrow{\kappa}\cdot\nabla T)$ along the magnetic field lines, using the Spizer conductivity $\kappa_{\parallel}=8\times 10^{-7}\ T^{5/2}\ \mathrm{ergs\ cm^{-1} s^{-1} K^{-1}}$ \citep{spitzer2006physics}. The comparably small value of $\kappa_{\perp}$ under solar coronal conditions allows us to neglect its contribution in this experiment.

The initial magnetic field is a force-free sheared arcade structure whose components are adopted as in \citet{Jenkins:2021aap} (see, Eqs. 8-10). However, we choose the bottom magnetic field strength to be $B_a=20$ G, and the parameter setting the lateral extension of the arcade to be $L_a=24$ Mm, in line with our physical box size.

We combine symmetric and antisymmetric boundary conditions at the side boundaries \citep[see their Table 1]{Jenkins:2021aap}. The velocities at the bottom ensure the converging and shearing motions ($V_x$ and $V_z$), while for $V_y$, we apply antisymmetry. The converging flow $V_x$ profile is defined following  Eq. 5 in \citet{Liakh:2020aap} assuming the parameters: $x_c=0$ Mm, $W=2L_a$, $\sigma=6.8$ Mm, while the shearing velocity is defined as $V_z=-V_x$. The temporal evolution of these imposed bottom boundary flows is defined by Eq. 6 in \citet{Liakh:2020aap} with the activation and deactivation times at $16.7\mins$ and $41.7\mins$, respectively. At the bottom, the density and pressure are fixed according to their initial values. The magnetic field $\mathbf{B_1}$ is set according to second-order zero-gradient extrapolation. The velocities and the magnetic field $\mathbf{B_1}$ at the top boundary are defined in accordance with third-order zero-gradient extrapolation. The density and pressure are set according to the gravitational stratification.

The randomized footpoint heating that imitates turbulent heating from the lower solar atmosphere is included after the first activation time ($t=16.7\mins$). Unlike previous studies by \citet{Zhou:2020nat} and \citet{Li:2022apj}, here we do not include the chromosphere and transition region. In order to be consistent with the previous experiments, we modify the parameter that controls the characteristic height that enters this random heating recipe to be $h_p=0\Mm$ (our bottom boundary, at low corona) in contrast to the value $3\Mm$ used by \citet{Li:2022apj}.

\section{Results}\label{sec:results}

 Following the mechanism proposed by \citet{vanBallegooijen:1989apj}, driving the arcade footpoints leads to (1) an increase in magnetic shear and (2) reconnection of strongly sheared field lines. The main novelty in our simulation is to account for the influence of randomized heating on FR formation. We point out that this random heating comes in multiple pulses that vary in their centroid locations, and that this makes the adopted heating spatio-temporally dependent, on top of the background heating $\mathcal{H}_b(y)$.
 
\begin{figure*}[!ht]
 	\begin{interactive}{animation}{Animation1.mp4}
 	\includegraphics[width=\textwidth]{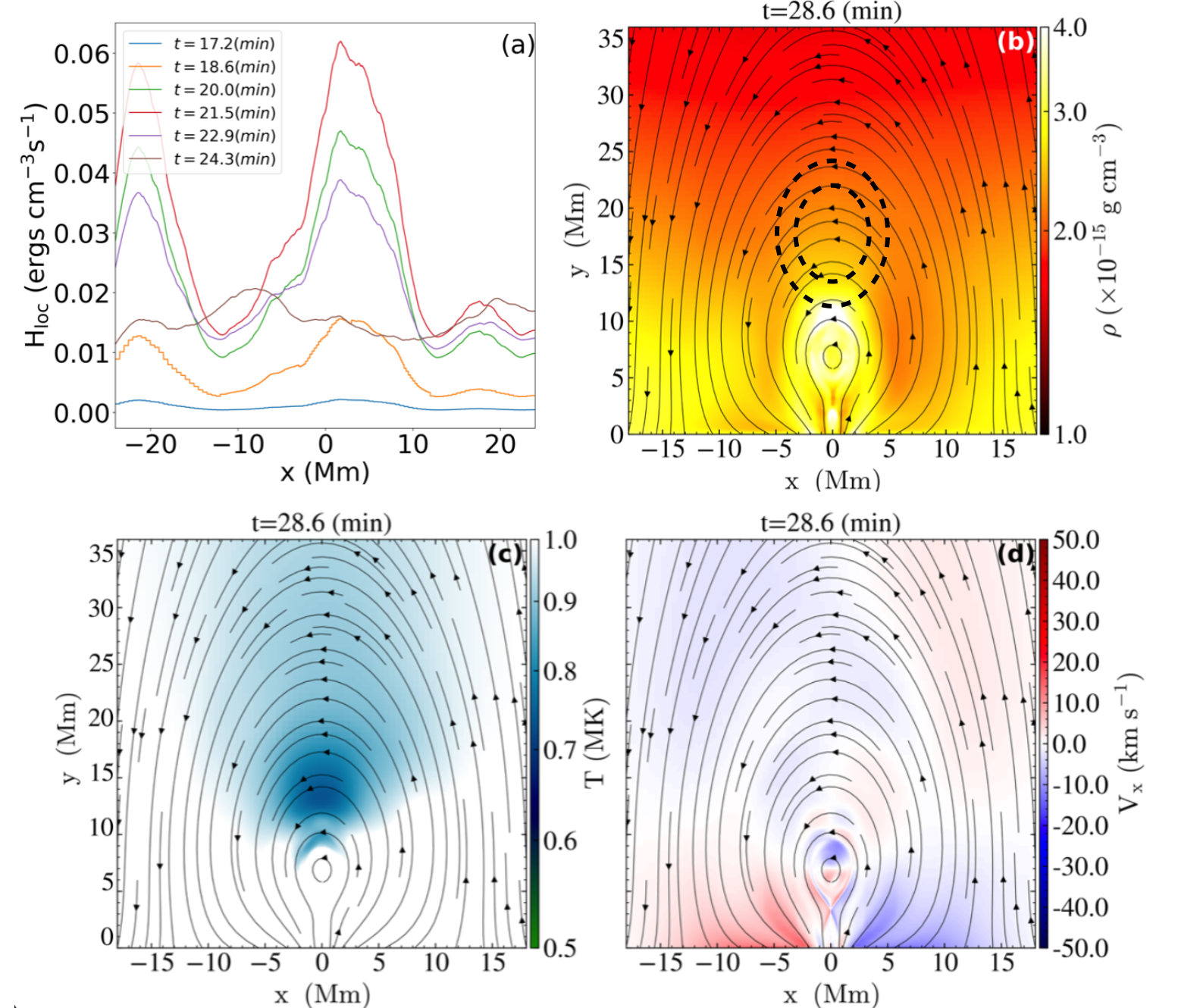}
 	 \end{interactive}
 	\caption{Localized heating (a) at our bottom corona at different times and a selected snapshot of the distribution of the density (b), temperature (c), and horizontal velocity (d) during the FR formation. The black lines denote the magnetic field lines in panels b-d. Two dashed black lines in panel (b) denote the position of the elliptical field lines used to further analyze plasma motions inside the flux rope. A movie of this figure is available with the online version of this manuscript (see Animation1.mp4). The movie shows the density (a),  temperature (b), and horizontal velocity (c) evolution in time-interval $17.2-142.5\mins$. The black lines denote the instantaneous magnetic field lines.
 		\label{fig:formation}}
 \end{figure*}

 Figure \ref{fig:formation}(a) shows that from $17.2$ to $21.5\mins$ (at about half the driving time period), the heating at the bottom increases, having a maximum value $0.06\ \mathrm{ergs\ cm^{-3}\ s^{-1}}$ at $x=0-10$ Mm. The heating distribution has another peak far from the reconnection site, of no impact to the FR formation. At $t=24.3\mins$, the heating profile changed to show a new peak at $x=-10\Mm$.

 The density distribution in panel (b) of Figure \ref{fig:formation} reveals that most material is collected at the top region of the FR and in overlying loops. This differs from previous studies by \citet{Kaneko:2015apj}, \citet{Kaneko:2015apj2}, \citet{Jenkins:2021aap}, and \citet{Brughmans:2022aap}, where lifted material located preferentially in the bottom FR region. This difference is due to the increased magnetic field strength $B_0=20$ G in contrast to $3-10$ G used previously. We analyze the magnetic tension, gas, and magnetic pressure forces along the vertical cut at $x=0\Mm$ and find significant magnetic tension above the forming FR caused by the stretching of field lines. The upward-acting magnetic and gas pressure forces also increase to balance the downward-acting magnetic tension, leading to compression at the top of the FR.

 Panel  (c) shows that the temperature decreases to approximately $0.5$ MK in the upper FR region and above, while the FR core is heated up to $2.5$ MK due to Ohmic dissipation. As a result of the randomized heating, the temperature distribution shows an asymmetry with respect to the polarity-inversion line (PIL) at $x=0\Mm$. The instantaneous horizontal velocity in panel (d) of Figure \ref{fig:formation} shows counterclockwise rotation right after the primary flux rope is formed.

The accompanying movie (Animation1.mp4) provides an overview of the entire experiment. 
Secondary FRs are formed at approximately $30$ and $36\mins$. Since the post-reconnection loops are already formed at $26\mins$, but the driving at the bottom is in action until $41.7\mins$, the footpoints of the post-reconnection loops are affected by the shearing and converging motions. As a consequence the post-reconnection loops form secondary FRs, which merge with the main FR.
 During this confluence, the flow patterns are more erratic. However, at $39\mins$, we can again observe clear rotation. These merging flux bundles supply plasma to the main FR \citep{Zhao:2022apj}.

The density and temperature distribution show that condensations start to occur at $40\mins$ when thermal instability (TI) develops. While forming, the condensed prominence takes part in the overall systematic rotation. After $t=50\mins$, we distinguish two prominence regions having different dynamics. The cold and dense prominence plasma close to the FR center rotates for the entire $142\mins$, while at a larger distance from the FR center, the condensations form, drop down and oscillate around their central magnetic dips. The evolution resembles the scenario proposed by  \citet{Wang:2010apjl} to explain the triggering of the observed tornadoes. We obtain counterclockwise spinning when the flow in the loops preceding the formation of the flux rope is directed from right to left.

\begin{figure*}[!ht]
	\includegraphics[width=0.5\textwidth]{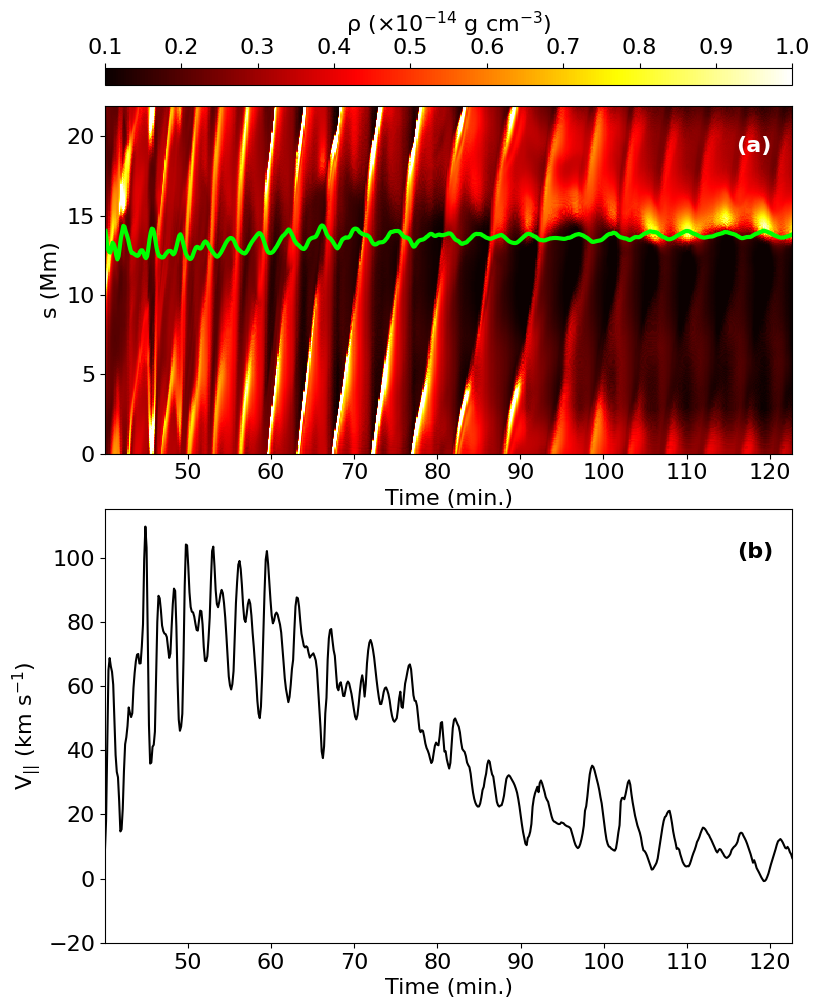}
	\includegraphics[width=0.5\textwidth]{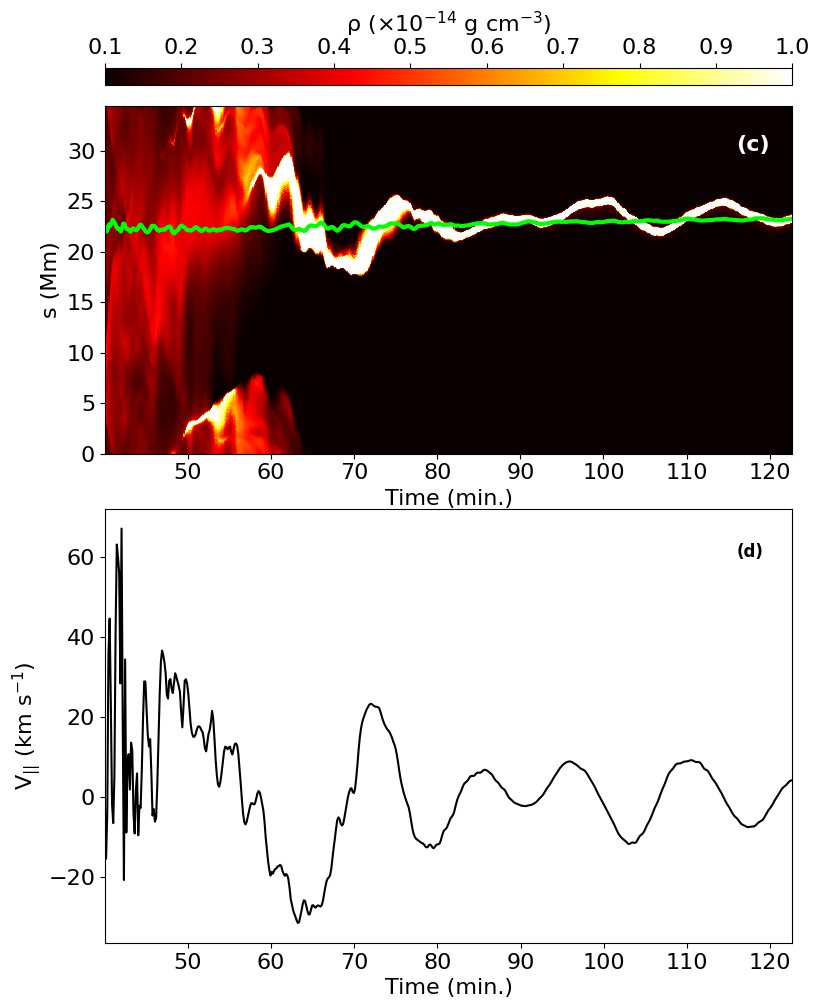}
	\caption{Time-distance diagram of the density along selected field lines (top) and poloidal field projected $v_{\parallel}$ at the center of mass of the same field lines (bottom). Left: the temporal evolution for the field lines with the dip at the height $y=10.8\Mm$ and radius of curvature $R_c=5.2\Mm$. Right: for the field line with dip at height $y=12.8\Mm$ and radius of curvature $R_c=3.0\Mm$. The green lines in panels (a)-(c) denote the position of the corresponding magnetic dip. \label{fig:time-distance}}
\end{figure*}
As the FR remains relatively stable, the plasma motions inside it can be studied along selected magnetic field lines as described by \cite{Liakh:2020aap}. To compare evolution closer to the center with that near the edge of the FR, we select two nested poloidal field lines with their radius of curvature $R_c=3.0\Mm$ and $5.2\Mm$ for the inner versus outer one, respectively.

Panel (a) of Figure \ref{fig:time-distance} shows the time-distance view of density, displaying plasma motions along the near-circular poloidal field lines. Rotation is then seen as repetitive bright ridges in plasma density (the flux-aligned $s$ parameter is periodic), with slopes that decrease up to time $90\mins$, suggesting the slowdown of the rotation. The velocity of the center of mass on that flux surface as shown in panel (b) shows that initially, the plasma rotates as fast as $60-95\kms$. However, after $60\mins$, these motions are reduced, and the field-projected poloidal velocity $v_{\parallel}=(v_xB_x+v_yB_y)/B$ decreases to $10-20\kms$.

 \citet{Li:2012apjl} studied a giant tornado observed by Solar Dynamic Observatory Atmospheric Imaging Assembly SDO/AIA in a large cavity of observed size $110-130\Mm$.
  Although the tornado observed by \citet{Li:2012apjl} was an activated mature prominence rather than the forming one, we can still qualitatively compare its evolution with our experiment when the flux rope is fully formed, and the swirling motions inside the cavity are established.
 This event had two phases, the first marking the onset of swirling motions from 2011 September 25, from 08:00 to 10:00. In the second phase, which lasted for $8$ hours, a significant mass supply from the chromosphere was observed. The authors inferred rotational velocities only for the second phase using local correlation tracking and obtained a value in the range of $55-95\kms$, in good agreement with our simulated velocities. \citet{Li:2012apjl} also estimated the radius of the circular motions in the second phase, around $90\Mm$, which is larger than our domain and all curvature radii in the flux rope. However, in the first observed phase before the mass supply, the rotating region was indeed much smaller. We also perform another numerical experiment with lateral extension, $L_a=96\Mm$, and magnetic scale height ,$H_b=200\Mm$. The rotating dynamics are very similar, but the size of the obtained tornado is larger than $20 \Mm$. Additionally, to have a mass supply similar to observations by \citet{Li:2012apjl}, including the chromosphere is needed. This can be addressed in a future study.

Panels (c) and (d) of Figure \ref{fig:time-distance} display plasma evolution associated with the field line with a larger radius of curvature. In this outer FR region, condensations form slightly later at $50-60\mins$ and show oscillatory motion around the center of the magnetic dip (this center is shown by the green line). This reveals a self-consistent triggering of oscillations together with the rotational flow.  The center of mass velocity $v_{\parallel}$ shows these oscillations reach maximum amplitude around $V_0=30\kms$ that allows us to classify them as large-amplitude oscillations (LAOs). Similarly semi-periodic motions are also present in the bottom region of the cavity 2011 September 25 between 11:00 and 12:00.

Using the Lomb-Scargle periodogram \citep{Lomb:1976apss}, we obtain the period of oscillations equal to $P=13 \mins$. Assuming the main restoring force is gravity projected along the magnetic field lines as suggested by many works \citep{Luna:2012apjl,Luna:2016aap,Zhang:2018apj,Zhang:2019apj,Liakh:2020aap,Fan:2020apj, Liakh:2021aap}, we can define the pendulum period from $P_{pend}=2\pi\sqrt{R_c/g_{\odot}}$, where $R_c=5.2\Mm$. This pendulum period is equal to $13.4\mins$ and is in excellent agreement.

When we fit the flow variation in panel (d) with a damped sinusoid function $v_{\parallel}=V_{0}e^{-t/\tau_{d}}\sin(2\pi t/P + \phi)$, we obtain the damping time equal to $\tau_D=40\mins$. There are several mechanisms that could account for the significant attenuation, including radiative losses and thermal conduction \citep{Zhang:2012aap,Zhang:2019apj}, mass accretion, particularly during condensation formation \citep{Ruderman:2016aap}, wave leakage \citep{Zhang:2019apj,Liakh:2021aap}, and energy exchange between the dips of different field lines \citep{Liakh:2021aap}. Additionally, one should assess the role of numerical dissipation \citep{Terradas:2016apj,Adrover:2020aap,Fan:2020apj,Liakh:2020aap}.


\begin{figure*}[!ht]
	\begin{interactive}{animation}{Animation2.mp4}
	\includegraphics[width=0.9\textwidth]{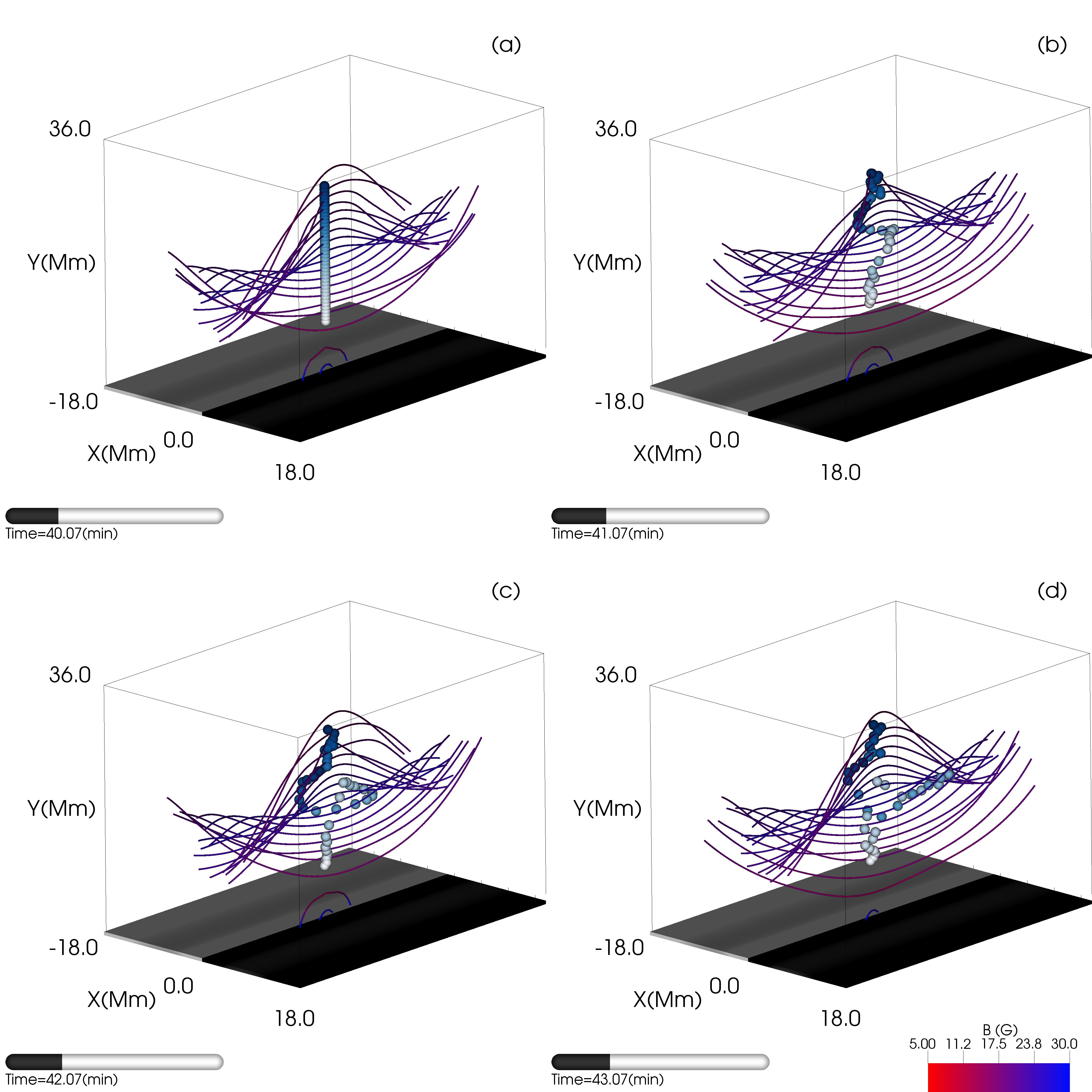}
	 \end{interactive}
	\caption{Temporal evolution of the magnetic field and corks in a 3D representation when the FR forms, and the rotation is established but prior to the condensation formation. The field line color denotes the magnetic field strength. The color of the corks corresponds to their different initial heights in the range of $7.5-27.5\Mm$. A movie of this figure is available with the online version of this manuscript (see Animation2.mp4). The movie shows the temporal evolution of the magnetic field and corks in a 3D representation in time-interval $40-45\mins$.
		\label{fig:3d-corks}}
\end{figure*}
So far, we have not considered the out-of-plane velocity component $v_z$. Our experiment reveals that $v_z$ can reach amplitudes of up to $100\kms$. To visualize the actual three-dimensional flow field, we use corks which allows us to obtain the 3D trajectories. Additionally, we visualize field lines using all three magnetic field components. Panel (a) of Figure \ref{fig:3d-corks} shows the initial positions of the selected corks and panel (b) demonstrates their displacement one minute after. In panels (c) and (d), taken at minute-interval we observe that the corks in the FR center show mainly axial motion, and slightly further, the corks exhibit predominantly rotation. The accompanying Animation2.mp4 indicates that most of the corks show an axial displacement in the same direction. However, there are several corks at the heights $10.5-12.5$ and $19-22\Mm$ that show slight shifts in the opposite direction. This 3D view hence shows that there is no significant PIL-aligned counterstreaming flow pattern, which is an alternative to explain the -- then apparent -- observed spinning motions.

\citet{Panasenco:2014solphys} explained rotation inside the cavity as 3D plasma motion along twisted magnetic fields. Our model is consistent with this suggestion. The 3D visualization in Figure \ref{fig:3d-corks} reveals that the evolution of the plasma shows helical motions inside the twisted FR. In future, more advanced 3D models featuring footpoint-anchored FRs, these helical plasma motions may result in mass drainage at one of the footpoints as described, for instance, in the observation analyzed by \citet{Wang:2017apj}.


\begin{figure*}[!ht]
	\begin{interactive}{animation}{Animation3.mp4}
	\includegraphics[width=0.9\textwidth]{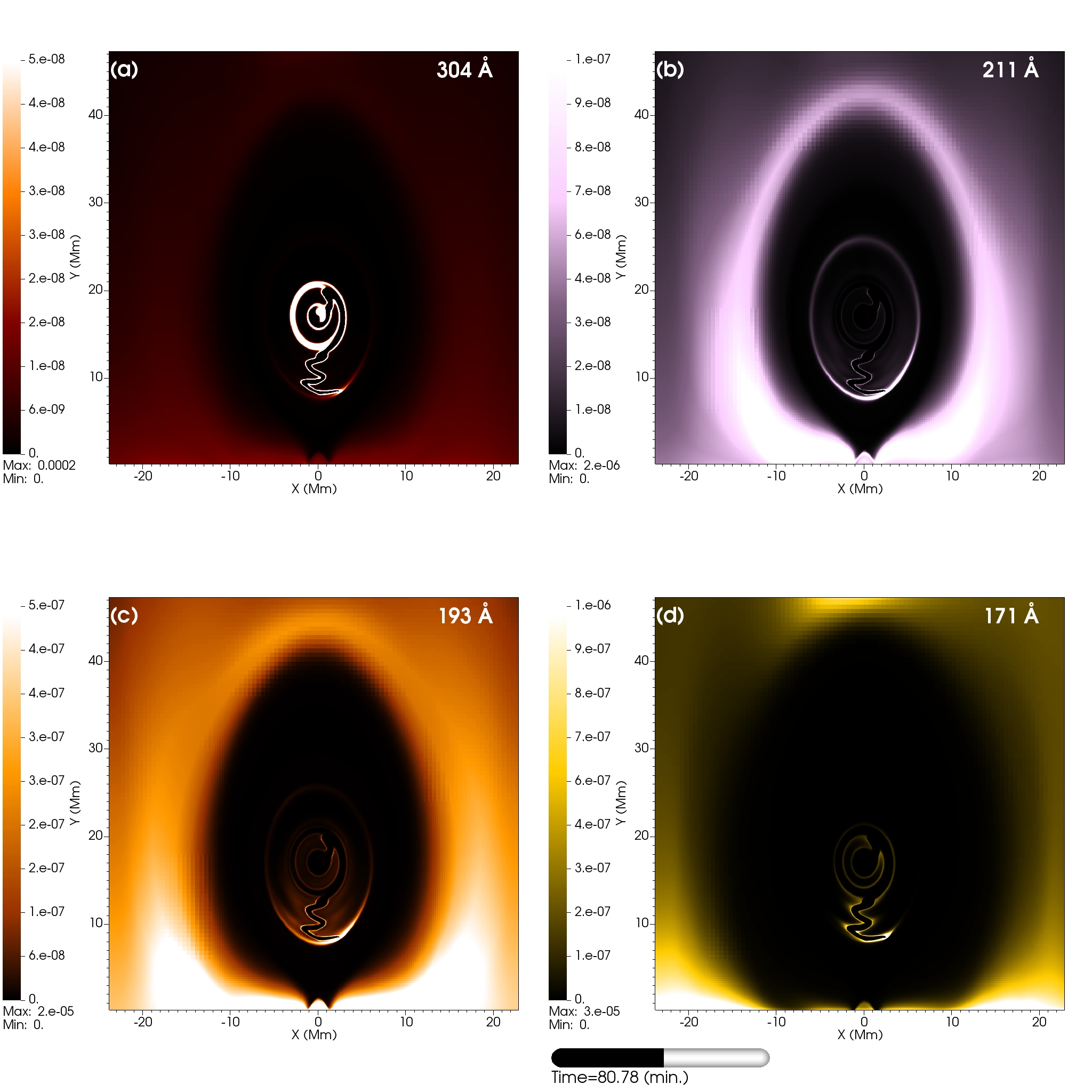}
    \end{interactive}
	\caption{Synthesized images in AIA channels: 304, 211, 193, and 171 \AA\, here shown at about $81\mins$. A movie of this figure is available with the online version of this manuscript (see Animation3.mp4). The movie shows the evolution during the FR formation and the rotational evolution ($16-141\mins$). 
		   \label{fig:euv}}
\end{figure*}
Following the method described by \citet{Xia:2014apjl}, we obtained synthesized images in four AIA channels 304 \AA\ (mostly temperatures at $10^4$ K), 211 \AA\  ($1.8$ MK), 193 \AA\ ($1.5$ MK), and 171 \AA\  ($0.8$ MK) which are shown in Figure \ref{fig:euv} at time $80.8\mins$. An accompanying movie (Animation3.mp4) covers a time interval of $16.2-140.8\mins$.
At $30.2\mins$, FR starts forming and growing in size, appearing as a bright structure in channels 211 and 193 \AA. The emission is caused by the coronal material inside the FR, which is denser than the surrounding corona. This bright structure suggests rotational behavior as it evolves to $36\mins$. At around this time, there is the first significant emission in channel 304 \AA\, inside the FR region, where the temperature decreases due to runaway cooling.

From $40.0$ to $58.2\mins$, the condensation forms and grows gradually. Due to the low temperature of the condensation, it appears as a dark structure in all four channels. In channel 304 \AA, only the prominence-corona transition region (PCTR) appears as a bright structure. It should be noted that this is a result of the approximate, optically thin treatment of 304 \AA\ in the synthetic images as explained by \citet{Xia:2014apjl}. Nevertheless, the prominence appearance in the different channels clearly manifests rotation along with the bright coronal plasma (see, panels (a-c) of Animation3.mp4).  Using SDO/AIA observations, \cite{Li:2012apjl} noticed an extremely similar appearance of their tornado in 304 and 171 \AA\ channels and also suggested that the cavity contained both hot and cold plasma.

At $58.2\mins$, the formation of a dark cavity can be seen in the 211 and 193 \AA\ channels. Initially, the darkening is caused by the strong cooling of this region due to the developing TI. Later, the FR recovers its temperature but still appears dark in the 211 and 193 \AA\ channels because it is strongly depleted of plasma due to the formation of condensations. Similarly, \citet{Li:2012apjl} showed the cavity as a dark structure best visible in 211 and 193 \AA\ channels from SDO/AIA observation. When the tornado developed, the authors observed bright plasma moving along the dark cavity, highlighting the magnetic field structure that was otherwise invisible. At the same time, channel 304 \AA\ displays the ongoing rotation of the prominence mostly seen through the bright PCTR inside the dark cavity. 

From $70.8\mins$, all four channels shows the dimming region in the loops surrounding the FR. As seen in Animation1.mp4, this region is also depleted of plasma. The border of the dimming region appears more luminous due to the compression of plasma in the overlying loops. At $102.5\mins$, the compressed material at the height of $50-60\Mm$ also condenses and drops down along the field lines.

\begin{figure*}[!ht]
	\includegraphics[width=0.99\textwidth]{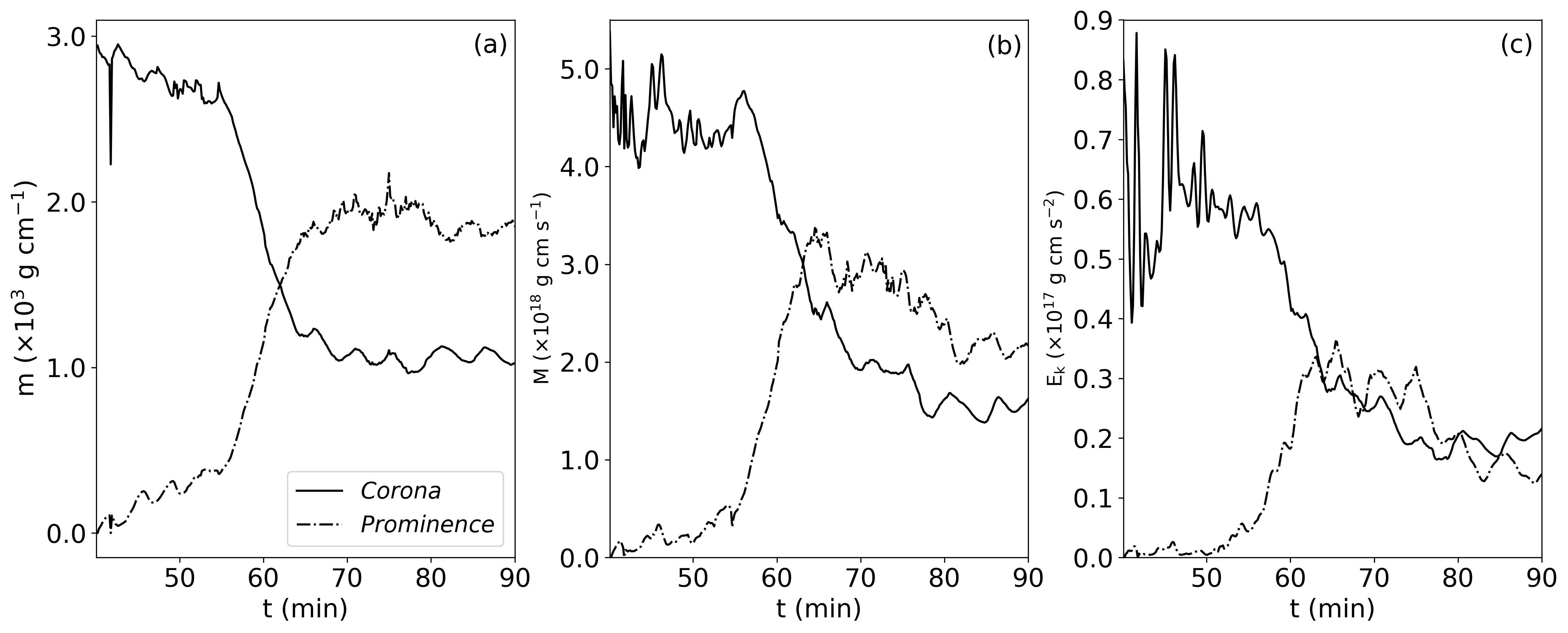}
	\caption{Temporal evolution of the total mass, angular momentum and kinetic energy of the coronal and prominence plasma in the dynamically evolving and tracked FR region.\label{fig:conservation}}
\end{figure*}
As seen from Animation1.mp4 and Animation3.mp4, from about $40$ to $65\mins$, both the coronal and prominence plasmas are involved in the rotations. We can then analyse the conserved quantities, such as total mass, angular momentum, and quantify kinetic energy changes during the condensation process. The FR region is therefore dynamically defined at each time iteration using the minimum and maximum values of the magnetic field curvature as described by \citet{Brughmans:2022aap}. Inside the defined FR region, we use a density threshold $\rho=10^{-14}\ \mathrm{g\ cm^{-3}}$  to detect the coronal and prominence plasma. We then compute total mass, angular momentum, and kinetic energy in the FR region for dense and tenuous plasma.

In panel (a) of Figure \ref{fig:conservation}, we observe that initially, FR contains only coronal plasma, as the condensation is not formed yet. From $40$ to $65\mins$, the mass, angular momentum, and kinetic energy of the prominence plasma rapidly increase when the condensation gradually forms, collecting more coronal material. In panel (b), the prominence angular momentum reaches its maximum at $65\mins$, when the total mass reaches is largest value $\sim 1.9\times 10^{3}\ \mathrm{g\ cm^{-1}}$. At the same time, the total mass of the coronal plasma contained inside FR is around $\sim 10^3\ \mathrm{g\ cm^{-1}}$, suggesting a decrease around $66\ \%$ in comparison to its value at $40\mins$. The clearly opposite trend between coronal and prominence matter in terms of mass and angular momentum content is fully consistent with the expected overall conservation of total mass and angular momentum within the FR.

After $65\mins$, the coronal and prominence mass remains nearly constant since no more condensations are formed. At the same time, angular momentum and kinetic energy of both coronal and prominence plasma show a decreasing trend. This decrease in total angular momentum and kinetic energy of both coronal and prominence plasma signals the gradual attenuation of motions. This also relates to the damping inferred for the outer oscillatory motions, and is due to the combination of non-adiabatic effects, our open top boundary, or numerical and actual resistive dissipation.

\section{Summary and conclusions}\label{sec:summary}

In this 2.5D numerical study, we model self-consistently the formation of giant tornadoes in their solar coronal cavity. This starts from a gravitationally stratified corona with a sheared arcade magnetic structure, subjecting it to footpoint converging and shearing to form an FR. Simultaneously with the FR formation, randomized heating at the bottom of the domain mimics the turbulent heating from the underlying solar atmosphere. The main findings of this work can be summarized as follows:
\begin{enumerate}
	\item The randomized heating results in the asymmetry of the temperature and density distributions with respect to the PIL, which leads to a coherent flow along the loops. This flow turns into a net rotational pattern as a twisted FR is formed.
	\item The condensed plasma in the central part of FR continues to rotate for over one hour, and the initial rotational velocity exceeds $60\kms$. Prominence plasma further from the FR center forms at the top, drains down and oscillates around the bottom of the magnetic dips with a velocity amplitude of around $30\kms$.
	\item The poloidal rotational plasma motions are visualized in 3D with corks trajectories which do not show significant counterstreaming flows, but suggest helical motions aware of the twisted magnetic field.
	\item The synthesized images in four AIA channels confirm the simultaneous rotation of the coronal plasma and the condensations. The formation of the dark cavity and dimming region in the overlying loops is also evident in 211 and 193 \AA\ images. 
\end{enumerate}

 The numerical model presented in this Letter finally explains and successfully reproduces the spinning motions observed inside coronal cavities. We plan to extend the model to a full 3D configuration allowing to see these rotating structures from different angles. Our model seriously challenges the observationally preferred interpretation where projection effects can disguise motions into apparent rotations \citep{Gunar2023}, making a 3D follow-up simulation urgently needed. Such 3D model could then reconcile this study with a recent analysis of a 3D prominence-cavity setup, concentrating on the ringlike, observed nested line-of-sight flow patterns without rotation \citep{Liu2023}.

\begin{acknowledgments}
\nolinenumbers We acknowledge support by the ERC Advanced Grant PROMINENT from the European Research Council (ERC) under the European Union’s Horizon 2020 research and innovation programme (grant agreement No. 833251 PROMINENT ERC-ADG 2018).
RK acknowledges support by the C1 project TRACESpace funded by KU Leuven and a FWO project G0B4521N. The computational resources and services used in this work were provided by the VSC (Flemish Supercomputer Center), funded by the Research Foundation Flanders (FWO) and the Flemish Government - department EWI. RK acknowledges the International Space Science Institute (ISSI) in Bern, ISSI international team project \#545.
\end{acknowledgments}

\bibliographystyle{aasjournal}
\bibliography{bibtex.bib}{}

\end{document}